\documentstyle[12pt]{article}
\begin{document}


\def\b #1{\mbox{\footnotesize\boldmath$#1$\normalsize}}

\def\bb#1{\mbox{\footnotesize\boldmath$\overline{#1}$\normalsize}}

\def\DG{\mbox{\large$\delta$\normalsize}}

\title{Is the third coefficient of the Jones knot polynomial a quantum
state of gravity?}

\author{\small Jorge Griego \\
\small Instituto de F\'{\i}sica, Facultad de Ciencias,\\
\small Trist\'an Narvaja 1674, Montevideo, Uruguay.\\
\small E-mail: griego@fisica.edu.uy}

\date{October 13, 1995}
\maketitle
\abstract{Some time ago it was conjectured that the coefficients of an
expansion of the Jones polynomial in terms of the cosmological constant
could provide an infinite string of knot invariants that are solutions
of the vacuum Hamiltonian constraint of quantum gravity in the loop
representation. Here we discuss the status of this conjecture at third
order in the cosmological constant. The calculation is performed in the
extended loop representation, a generalization of the loop
representation. It is shown that the the Hamiltonian does not annihilate
the third coefficient of the Jones polynomal ($J_3$) for general
extended loops. For ordinary loops the result acquires an interesting
geometrical meaning and new possibilities appear for $J_3$ to represent
a quantum state of gravity.}

\section{Introduction}
One of the most promising achievements of the new variable canonical
cuantization program of gravity \cite{As} is the possibility to find
in a generic case solutions to the Wheeler-DeWitt equation. This fact
has allowed to remove one of the main difficulties that stoped the
hamiltonian cuantization of gravity for almost thirty years. If the
space of states could be determined in any appropriate way (that is, in
any well defined formalism associated with gravity), then one has the
chance to advance one more step in the long avenue of the quantum theory
of gravity. In the case of the Ashtekar formalism this avenue present
special features (as the fact that the variables are complex and reality
conditions need to be imposed at the quantum level) that could reveal
unexpected difficulties to complete the whole program. But all future
advance necessarily goes throught the knowledge of the physical space of
states of the theory, that is, of the solutions of the diffeomorphism
and the Hamiltonian constraints.

There are mainly two ingredients of the new program that guide the
search of solutions of the constraints of quantum gravity: the loop
representation and its connection with knot theory \cite{RoSm} and the
relationship between Chern-Simons theory and the Kauffman bracket and
Jones knot polynomials \cite{BrGaPu1}. The existence of a loop
representation for quantum gravity is a direct consequence of the new
variables introduced by Ashtekar (connections and triads  instead of
metrics and conjugated momenta). Whenever one has a theory given in
terms of a Lie-algebra valued connection on a three manifold, one can
introduce a loop representation for it \cite{GaTr}. The fundamental goal
of the loop representation is that it has allowed to find for the first
time nonperturbative solutions of the Wheeler-De Witt equation.

The loop representation has an intrinsic geometrical content that
simplifies notably the constraint equations. The invariance under
diffeomorphism of the theory can be immediately coded in the requirement
of knot invariance. This fact automatically solves the diffeomorphism
constraint, so we have to deal here only with the Hamiltonian constraint.

One can take different points of view for the analysis of the
Hamiltonian constraint in the loop representation. One possible
approach is to consider the geometrical properties of loops rather
than the explicit analytical expressions of the knot invariants. This
point of view was the first adopted historically and it was based
initially on the following observation: the action of the Hamiltonian is
automatically zero over {\it smooth} loop wavefunctions (that is, over
wavefunctions that are nonzero only for loops without kinks and
intersections). Using this fact, it is possible to give a
prescription that connects the space-time metric with some underlying
structure constructed in terms of smoothened loops (the weaves)
\cite{AsRoSm}.

Another possible approach to the problem is by using the analytical
expression of the knot invariants to evaluate the Hamiltonian.
This apparently trivial observation is in fact amazing. We know only a
few analytical expressions  for the so many knot invariants that one can
construct in knot theory. On the other hand, any explicit calculation of
the Hamiltonian over a loop wavefunction implies to compute the loop
derivative \cite{Ga} over some kind of intersection and this means to
handle with hard computational and regularization problems. In fact, the
only nondegenerate solution of the vacuum Hamiltonian constraint found
up to now by this method corresponds to the second coefficient of the
Alexander-Conway knot polynomial and the result is only formal
\cite{BrGaPu2}. In spite of this, there exists a guideline that increase
the expectative to find a systematic method to generate solutions of the
Hamiltonian constraint. This is the content of the second ingredient
mentioned above, the relationship between Chern-Simons theory and the
Kauffman bracket and Jones knot polynomials. The possibility that these
knot polynomials might be associated with quantum states of gravity is
considered in the next section. The explicit exploration of this fact
faces up with very nontrivial technical difficulties in the loop
representation.

Recently a representation closely related to the loop representation was
introduced for quantum gravity \cite{DiGaGrPu1,DiGaGr2}. This
representation is based on an extension of the group of loops to a local
infinite dimensional Lie group \cite{DiGaGr1} and it is called by this
reason the extended loop representation. In spite that some problems
appear at the basic level of the extended loop representation (the
convergence problems associated with the matrix representations of the
extended holonomies and its relationship with the gauge invariance
properties of the representation \cite{Tr}), the extended loop formalism
was able to show an interesting ability to handle the regularization and
renormalization problems and to increase the power of calculus of the
theory. These and other related questions are discussed in reference
\cite{DiGaGrPu2}.

The aim of this paper is to exhibit the computational power of this new
representation and show that relevant information about the states of
quantum gravity can  be obtained. In particular we will show under which
conditions the third coefficient of  a certain expansion of the Jones
polynomial can be viewed as a solution of the Wheeler- DeWitt equation.
The main result is that, when restricted to ordinary loops the third
coefficient could be in principle annihilated by the Hamiltonian for
some particular topologies of the loop at the intersections. Moreover,
it is shown that only for such privileged topologies the $J_3$ knot
invariant would satisfy all the Mandelstam identities required
for the quantum states of gravity in the loop representation.

The article is organized as follows: in Sect. 2 the loop transform of
the exponential of the Chern-Simons form is considered. This state is
the key that allows to relate solutions of the constraints of gravity in
the connection and the loop representations. In Sect. 3 a result for
$H_0\,J_3$ is derived in terms of extended loops. This section includes
a subsection where the tools of the extended loop framework are
introduced. The reduction of the result from extended to ordinary loops
is performed in Sect. 4. The implications of the procedure of reduction
for quantum gravity are considered and, in particular, the Mandelstam
identities of $J_3$ are discussed in Sect. 4.1. The conclusions are
included in Sect. 5 and two appendixes with some useful derivations are
added.

\section{The exponential of the Chern-Simons \protect\\form}

Kodama \cite{Ko} was the first to recognize that the exponential of the
Chern-Simons form $\Psi_{CS}[A]$ constructed with the Ashtekar
connection gives an exact quantum state of gravity with cosmological
constant $\Lambda$:

\begin{equation}
H_{\Lambda}[A]\,\Psi_{CS}[A]=
(H_{0}[A]+\textstyle{\Lambda\over6}\,det\,q[A])\,\Psi_{CS}[A]=0
\label{hcs}
\end{equation}
$H_{0}[A]$ and $det\,q[A]$ are the Hamiltonian and the determinant of
the three metric in the connection representation. The connection and
the loop representations are related through the (formal) loop transform
\begin{equation}
\Psi[\gamma]=\int dA \,W_{\gamma}[A]\, \Psi[A]
\end{equation}
where $W_{\gamma}[A]:=Tr[U_A(\gamma)]$ is the Wilson loop and
$U_A(\gamma):=P\,exp(\int_{\gamma}A_a(x)dx^a)$ is the holonomy. At
least formally one can consider the loop transform of $\Psi_{CS}(A)$:
\begin{equation}
\Psi_{CS}[\gamma]=\int dA \,W_{\gamma}[A]\, e^{-\textstyle{6
\over \Lambda} S_{CS}(A)}\;\,.
\label{lpcs}
\end{equation}
This expression looks the same as the expectation value of the Wilson
loop in a Chern-Simons theory and has been studied by many authors
\cite{Wi,Sm,Gumami}. The result is that it is a knot invariant that is
known as the Kauffman bracket knot polynomial. The loop transform of
equation (\ref{hcs}) promotes then, at least formally, the
Kauffman bracket as a solution of the Hamiltonian constraint with
cosmological constant in the loop representation. This fact was
confirmed by Br\"ugmann, Gambini and Pullin \cite{BrGaPu1} up to the
second order in the cosmological constant. The extended loop version of
these calculations has showed that the result is also valid into a
regularized and renormalized context \cite{DiGaGr2,Gr2}

The Kauffman bracket and the Jones polynomial are related through the
following expression

\begin{equation}
K_{\Lambda}(\gamma)= e^{-\Lambda \varphi_G (\gamma)}\,J_{\Lambda}(\gamma)
\end{equation}
where $\varphi_G (\gamma)$ is the Gauss self-linking number of $\gamma$
and we have rescaled the cosmological constant
($\Lambda_{\mbox{\scriptsize{new}}}\equiv\Lambda_{\mbox{\scriptsize{old}}}/6$).
Let $K_{\Lambda}(\gamma)= \sum_{m=0}^{\infty} \Lambda^m K_m (\gamma)$
and $J_{\Lambda}(\gamma)= \sum_{n=0}^{\infty} \Lambda^n J_n (\gamma)$ be
the expansions of the Kauffman and Jones polynomial in terms of the
cosmological constant, then

\begin{eqnarray}
H_{\Lambda} \, K_{\Lambda}(\gamma) &=& \sum_{m=0}^{\infty}
\Lambda^m\,(H_{0}+\Lambda\,det\,q)K_m (\gamma)\nonumber\\
&=&\sum_{m=1}^{\infty}
\Lambda^m\,[H_{0}\,K_m (\gamma)+det\,q\,K_{m-1}(\gamma)]
\label{hkauff1}
\end{eqnarray}
with
\begin{equation}
K_m =\sum_{n=0}^{m}
{\textstyle{\frac{(-1)^n}{n!}}}
\,\varphi^n_G \,J_{m-n}
\label{km}
\end{equation}
The exponential of the Gauss self linking number is by itself a solution
of the Hamiltonian constraint with cosmological constant
\cite{GaPu,Gr2}. This means that

\begin{equation}
H_{\Lambda} \, e^{-\Lambda \varphi_G (\gamma)} = \sum_{n=0}^{\infty}
\Lambda^n\,\textstyle{\frac{(-1)^n}{n!}}
\,[H_{0}\,\varphi^n_G -n\,det\,q\,\varphi^{n-1}_G]=0
\end{equation}
Using this fact it is easy to show that (\ref{hkauff1}) can be put in
the form
\begin{eqnarray}
&&\hspace{-0.7cm}
H_{\Lambda} \, K_{\Lambda}(\gamma) = \nonumber\\
&&\hspace{-0.7cm}\sum_{m=2}^{\infty}
\Lambda^m \!\{ H_0\,J_m+\sum_{n=1}^{m-1}
{\textstyle{\frac{(-1)^n}{n!}}}
[H_0 \,(\varphi^{n}_G \,J_{m-n})-n\,det\,q \,(\varphi^{n-1}_G \,J_{m-n})]\}
\label{hkauff}
\end{eqnarray}
If the Kauffman bracket is annihilated by $H_{\Lambda}$, each term of the
sum in $m$ has to vanish for separate. For $m=2$ the above result reduces to
(recall that $J_1\equiv 0$)

\begin{equation}
H_0\,J_2 (\gamma)=0
\end{equation}
$J_2 (\gamma)$ coincides with the second coefficient of the
Alexander-Conway polynomial, so this result is the same that the
obtained by Br\"ugmann, Gambini and Pullin \cite{BrGaPu2} through an
explicit computation. This fact aimed these authors to make the
conjecture that the same result could also hold for higher orders in
$\Lambda$. This means that some cancellation mechanism must operate in
order to make $H_0\, J_m (\gamma)=0$ for all $m$ (the cancellation of
the sum of square brackets in (\ref{hkauff})). If this is true, the
expansion of the Jones polynomial in terms of the cosmological constant
would provide an infinite string of knot invariants that are annihilated
by the vacuum Hamiltonian constraint.

To third order in $\Lambda$ equation (\ref{hkauff}) reads
\begin{equation}
H_{\Lambda} \, K^{(3)}_{\Lambda} =
\Lambda^3 \{ H_0\,J_3+det\,q \,J_2-H_0 \,(\varphi_G \,J_2)  \}
\label{hk}
\end{equation}
The analytical expressions of the knot invariants included in the above
equation are the followings:
\begin{eqnarray}
&&\varphi_G (\gamma)=\textstyle{3\over2} g_{\mu_1 \mu_2} X^{\mu_1
\mu_2}(\gamma) \label{gauss}\\
&&J_2 (\gamma)= -3\{h_{\mu_1 \mu_2 \mu_3} X^{\mu_1 \mu_2 \mu_3}(\gamma)+
g_{\mu_1 \mu_3}g_{\mu_2 \mu_4} X^{\mu_1 \mu_2 \mu_3\mu_4}(\gamma)\}
\label{J2}\\
&&J_3 (\gamma) = -6\{
(2g_{\mu_1\mu_4}g_{\mu_2\mu_5}g_{\mu_3\mu_6} +
{\textstyle{1 \over 2}}
g_{(\mu_1\mu_3}g_{\mu_2\mu_5}g_{\mu_4\mu_6)_c}) \,
X^{\mu_1\mu_2\mu_3\mu_4\mu_5\mu_6}(\gamma)
\nonumber \\
&&\hspace{3cm}+ g_{(\mu_1\mu_3}h_{\mu_2\mu_4\mu_5)_c} \,
X^{\mu_1\mu_2\mu_3\mu_4\mu_5}(\gamma)
\nonumber \\
&&\hspace{2.2cm}
(h_{\mu_1\mu_2\alpha} g^{\alpha\beta} h_{\mu_3\mu_4\beta} -
h_{\mu_1\mu_4\alpha} g^{\alpha\beta} h_{\mu_2\mu_3\beta}) \,
X^{\mu_1\mu_2\mu_3\mu_4}(\gamma)\}
\label{J3}
\end{eqnarray}
The loop dependence of the knot invariants are written in terms of the
multitangents fields $X^{\mu_1\ldots\mu_n}(\gamma)$, which are defined
as distributions integrated along the loop $\gamma$:

\begin{equation}
X^{\mu_1 \ldots \mu_n}(\gamma) :=
\oint_\gamma dy_n^{a_n} \! \ldots \! \oint_\gamma dy_{1}^{a_1} \delta
(x_n\!-y_n) \! \ldots \! \delta(x_1\!-y_1) \Theta_\gamma(o,y_1,\! \ldots
\! ,y_n)
\label{multitangent}
\end{equation}
The greek indices of the multitangents represent a pair of vector index
and space point ($\mu_i:=(a_i,x_i)$). The $\Theta$ function orders the
points of integration along $\gamma$ with origin $o$ and
$g_{\mu_1\mu_2}$ and $h_{\mu_1\mu_2\mu_3}$ are the two and three point
propagators of the Chern-Simons theory, given by

\begin{equation}
g_{\mu_1 \mu_2}   = - {\textstyle{1 \over4\pi}}\epsilon_{a_1 a_2 k}
\frac{(x_1 -x_2)^k}{\mid x_1 -x_2\mid^3}= - \epsilon_{a_1 a_2 k} \,
\frac{\partial^{\,k}}
{\nabla^{2}} \, \delta(x_1 -x_2)
\label{metrica}
\end{equation}
\begin{equation}
h_{\mu_1 \mu_2 \mu_3} = \epsilon^{\alpha_1 \alpha_2 \alpha_3} \;
g_{\mu_1 \alpha_1} \, g_{\mu_2 \alpha_2}\, g_{\mu_3 \alpha_3}
\label{h}
\end{equation}
with
\begin{equation}
\epsilon^{\alpha_1 \alpha_2 \alpha_3} = \epsilon^{c_1 c_2 c_3}
\int d^3 t \; \delta(z_1 -t)\,\delta(z_2 -t)\,\delta(z_3 -t)
\end{equation}
In equations (\ref{gauss})-(\ref{J3}) a generalized Einstein convention
is assumed (the repeated vector indices are summed from $1$ to $3$ and
the spatial variables are integrated in ${\cal R}^3$).

The above expressions admit a direct translation to the extended loop
space. In fact, extended loops were introduced as generalizations of
the multitangent fields to includ more general fields. These fields are
multivector densities that satisfy two conditions: the differential and
algebraic constraints, given by

\begin{equation}
\partial_{\mu_i} X^{\mu_1 \ldots \mu_i \ldots \mu_n} =
\left[ \delta(x_i-x_{i-1}) - \delta(x_i-x_{i+1}) \right]
 X^{ \mu_1 \ldots \mu_{i-1} \; \mu_{i+1} \ldots \mu_n }
\label{dc}
\end{equation}
and
\begin{equation}
X^{\underline{\mu_1\ldots\mu_k}\mu_{k+1}\ldots\mu_n} = X^{\mu_1\ldots\mu_k} \,
X^{\mu_{k+1}\ldots \mu_n}
\label{ac}
\end{equation}
In (\ref{ac}) the underline means a sum over all the permutations that
preserve the order of the $k$ and $n-k$ indices among themselves. The
differential constraint depends on a basepoint $o$ (we have
$x_0=x_{n+1}\equiv o$ in (\ref{dc})), that coincides with the origin of the
loops when $X^{\b \mu}=X^{\b \mu}(\gamma)$. The constraints are related
to properties of the multitangent fields: the differential constraint
(\ref{dc}) is associated with the behavoir of the holonomies
under gauge transformations and the algebraic constraint
(\ref{ac}) is a consequence of the existence of an order of
integration along the path (the $\Theta$ function that
appears in (\ref{multitangent})). In general we write the elements of the
extended loop group as infinite strings of the form
\begin{equation}
{\bf X} \, =   (\, X, \, X^{\, \mu_1}, \, \ldots , \,
X^{\,\mu_1\ldots \mu_n},\, \ldots )=(\, X, \, X^{\b \mu})
\end{equation}
where $X$ is a real number and $\b \mu=(\mu_1\ldots \mu_n)$ represents
a set of indices of rank $n(\b \mu)=n$ (the rank of a multivector is
given by the number of paired indices). The extended group product is
defined through the expression
\begin{equation}
({\bf X}_1 \times {\bf X}_2)^{\b \mu}   =
\DG^{\b \mu}_{\b \pi \b \theta} \, X_1^{\b \pi}\, X_2^{\b \theta}
\label{gpro2}
\end{equation}
where the matrix $\DG^{\b \alpha}_{\b \beta}$ is given by the following
product of discrete and continuous delta functions
\footnote{ $\delta^{\alpha_i}_{\beta_i}:=\delta^{c_i}_{d_i}\delta(z_i-y_i)$,
with $\alpha_i:=(c_i,z_i)$ and $\beta_i:=(d_i,y_i)$. }:
$\delta^{\alpha_1}_{\beta_1} \cdots \delta^{\alpha_n}_{\beta_n}$ if the
sets $\b \alpha$ and $\b \beta$ have equal rank and zero otherwise. In
(\ref{gpro2}) the sets $\b \pi$ and $\b \theta$ are summed from rank
zero to rank infinite. Notice that for $n(\b \pi)=0$ and $n(\b
\theta)=0$ one gets the components of rank zero of the multitensor
strings.

The Hamiltonian and the determinant of the three metric can be
implemented in the extended space. So one can verify if $H_{\Lambda}
\, K^{(3)}_{\Lambda} =0$ in the extended loop representation. The
evaluation of $H_0\,J_3({\bf X})$ is possible but implies a long
calculation. Here we adopt the following point of view: we accept that
the Kauffman bracket is annihilated by $H_{\Lambda}$ and then we derive
a result for $H_0\,J_3({\bf X})$ from the annulment of
(\ref{hk}). The confirmation of the fact that $H_{\Lambda} \,
K^{(3)}_{\Lambda} =0$ will be given elsewhere \cite{Gr}.

\section{A result for $H_0\,J_3({\bf X})$}
We start this section introducing the fundamental tools of the extended
loop framework.

\subsection{Extended loops and quantum gravity}
The extended loop representation of gravity is constructed in the group
${\cal D}_{o}$ whose elements satisfy only the differential constraint.
The wavefunctions are {\it linear} in the multivector fields and the
Mandelstam identities make them to depend on the following combination
of $X$'s

\begin{equation}
\psi({\bf   X})   \equiv \psi({\bf   R})=   D_{\b \mu}   \,    R^{\b \mu}
\end{equation}
with
\begin{equation}
R^{\b \mu} := \textstyle{1\over2} [  X^{\b  \mu}  +  X^{\bb \mu}]
\end{equation}
The overline indicates two operations: the reversal of the sequence of
indices and a sign that depends on the rank of the set; that is

\begin{equation}
X^{\bb \mu} := (-1)^{n(\b \mu)}
X^{\b \mu^{-1}}
\end{equation}
with $\b \mu^{-1}:=(\mu_n\ldots\mu_1)$. The expression of the
constraints in terms of extended loops are

\begin{equation}
{\cal  C}_{ax} \psi({\bf R})
= \psi({\cal  F}_{ab}(x) \times {\bf R}^{(bx)} )
\label{extdif}
\end{equation}
\begin{equation}
\textstyle{1\over2}H_0 (x) \, \psi({\bf R}) =
\psi ({\cal F}_{ab} (x) \times {\bf   R}^{(ax, \,bx)})
\label{ham}
\end{equation}
The action of the diffeomorphism (\ref{extdif}) and the Hamiltonian
(\ref{ham}) operators reduce to a shift in the argument of the
wavefunctions. The shifted arguments are given in both cases by the
group product between an element of the algebra (the ${\cal F}_{ab}(x)$)
and a combination of ${\bf R}$'s with one and two spatial points
evaluated at $x$. The ${\cal F}_{ab}(x)$ has only two nonvanishing
components

\begin{eqnarray}
&& {\cal F}_{ab}{^{a_1 x_1}}(x)   =
\delta_{a\,b}^{a_1 \!\;  d} \; \partial_d
\, \delta(x_1 - x) \\
&&  {\cal  F}_{ab}{^{a_1 x_1\,a_2 x_2}}(x)=\delta_{a\,b}^{a_1\, a_2}
\; \delta(x_1 - x) \, \delta(x_2 - x)\;\;,
\end{eqnarray}
and the ``one-point-R" and the ``two-point-R" are given by the following
combinations of multivector fields:

\begin{equation}
[{\bf R}^{(bx)}]^{\b \mu} \equiv R^{(bx) \b \mu}
:= R^{(bx \, \b \mu)_c}
\end{equation}
\begin{equation}
[{\bf   R}^{(ax, \,bx)}]^{\b \mu} \equiv
R^{ (ax, \,bx) \b \mu} :=
\DG_{\b \pi \b \theta}^{\b \mu}\,
R^{(ax\, \b \pi \,bx\,\bb \theta)_c}
\label{dos}
\end{equation}
The subscript $c$ indicates cyclic permutation. For the determinant of
the three metric one finds the result

\begin{equation}
\textstyle{1\over2}det \,q (x)\, \psi({\bf R}) =
\psi (\epsilon_{abc} \,{\bf   R}^{(ax,\, bx, \,cx)})
\label{detq}
\end{equation}
where the ``three-point-R" is given by the following expression

\begin{equation}
R^{ (ax,\, bx, \,cx) \b \mu} :=
\DG_{\b \alpha \b \beta \b \gamma}^{\b \mu}\,
[2\,R^{(ax\, \b \beta \,bx\, \b \alpha \,cx\,\bb \gamma)_c}+
R^{(ax\, \b \alpha \,bx\, \b \beta \,cx\,\bb \gamma)_c}]
\label{tres}
\end{equation}

The diffeomorphism and the Hamiltonian have very similar expressions
when they are written in terms of extended loops. The only difference is
the object that one puts into the group product with ${\cal F}_{ab}(x)$.
So, the difference of the results would depend on the properties of the
one- and two-point-R. The difference lies basically in their symmetry
and regularity properties. A multivector field with two indices
evaluated at the same spatial point generates a divergence. This is due
to the distributional character of the multitensors. A multitensor
satisfying the differential constraint (\ref{dc}) diverges when two
successive indices are evaluated at the same spatial point. This
divergence can be regularized introducing point-splitting smearing
functions. In spite of this, we develop here only the formal calculation
for the sake of simplicity. The regularization and renormalization of
the formal result for $H_0\,J_3$ involves several particular features
that will be given elsewhere.

Notice the effect of ${\cal F}_{ab}(x)$ in the general expression of
the constraints. This quantity has only two nonvanishing components of
rank one and two, so

\begin{equation}
{\textstyle{1\over2}}H_0 (x)  \psi({\bf R}) =
\sum^{\infty}_{n=0} D_{\mu_1 \ldots \mu_n}[
{\cal F}_{ab}^{\mu_1} (x)\,   R^{(ax,\,bx)\mu_2 \ldots \mu_n}+
{\cal F}_{ab}^{\mu_1 \mu_2} (x)\,   R^{(ax,\,bx)\mu_3 \ldots \mu_n}]
\label{ham2}
\end{equation}
Duo to the Mandelstam identities the propagators are  cyclic under
permutation of the indices ($D_{\mu_1 \ldots \mu_n}=D_{(\mu_1 \ldots
\mu_n)_c}$). This means that the indices of the propagator that are
contracted with ${\cal F}_{ab}(x)$ really lie in any position of the
$D$'s. In general the result of this contraction is to  modify the
structure of the propagator and to fix some indices of the propagator
and of the two-point-R at $x$.

\subsection{The calculation}

{}From the annulment of (\ref{hk}) one has
\begin{equation}
H_0\,J_3= H_0 \,(\varphi_G \,J_2)-det\,q \,J_2
\end{equation}
In order to compute the first contribution of the r.h.s. we
need to define the action of an operator onto the  product of two
invariants (remember that the operators only know to act on linear
expressions of the multivector fields). The linear wavefunction that
corresponds to the product of $\varphi_G$ and $J_2$ is taken to be
the following \footnote{The star product can be defined in a rigorous
manner in the extended framework and it satisfies several interesting
properties, see \cite{star}.}:

\begin{eqnarray}
\varphi_G ({\bf X})\,J_2 ({\bf X}) \rightarrow (\varphi_G * J_2) ({\bf X})
 &:=&-\textstyle{9\over2} \{g_{\mu_1 \mu_2}
h_{\mu_3 \mu_4 \mu_5} X^{\underline{\mu_1 \mu_2} \mu_3\mu_4\mu_5}
\nonumber\\
&&+g_{\mu_1 \mu_2}g_{\mu_3 \mu_5}g_{\mu_4 \mu_6}
X^{\underline{\mu_1 \mu_2} \mu_3\mu_4\mu_5\mu_6} \}
\end{eqnarray}
In ${\cal D}_o,\;(\varphi_G * J_2)$ defines the wavefunction ``product"
of $\varphi_G$ and $J_2$. For any multitensor that satisfy the algebraic
constraint the above expression reduces to the usual product of the two
diffeomorphism invariants. From (\ref{gauss}), (\ref{J2}) and
(\ref{ham2}) we then have

\begin{eqnarray}
\lefteqn{ H_0\,(\varphi_G * J_2) ({\bf R}) =
-9 [\,g_{\mu_1 \mu_2}
h_{\mu_3 \mu_4 \mu_5} ({\cal F}_{ab}(x)\times {\bf R}^{(ax,\,bx)})
^{\underline{\mu_1 \mu_2} \mu_3\mu_4\mu_5} }
\nonumber\\
&&+g_{\mu_1 \mu_2}g_{\mu_3 \mu_5}g_{\mu_4 \mu_6}
({\cal F}_{ab}(x)\times {\bf R}^{(ax,\,bx)})
^{\underline{\mu_1 \mu_2} \mu_3\mu_4\mu_5\mu_6} \,]
\label{hgj2}
\end{eqnarray}
It is possible to prove that \cite{Gae}
\begin{eqnarray}
&&({\cal F}_{ab}(x)\times
{\bf R}^{(ax,\,bx)})^{\underline{\mu_1 \ldots \mu_k}\mu_{k+1} \ldots \mu_n}=
\sum_{i=0}^{k}{\cal F}_{ab}^{\mu_1 \ldots \mu_i}(x)
R^{(ax,\,bx)\mu_{i+1} \ldots \mu_k \underline{\mu_{k+1} \ldots \mu_n}}
\nonumber\\
&&\hspace{1.313cm}+\sum_{i=k}^{n}{\cal F}_{ab}^{\mu_{k+1} \ldots \mu_i}(x)
R^{(ax,\,bx)\underline{\mu_1 \ldots \mu_k}\mu_{i+1} \ldots \mu_n}
\end{eqnarray}
This result follows from the fact that ${\cal F}_{ab}(x)$ satisfies the
homogeneous algebraic constraint. Developing (\ref{hgj2}) according to
this rule one obtains
\begin{eqnarray}
\lefteqn{ H_0\,(\varphi_G * J_2) ({\bf R}) =
-9 [\,g_{\mu_1 \mu_2}h_{\mu_3 \mu_4 \mu_5}
({\cal F}_{ab}(x)^{\mu_1}R^{(ax,\,bx)\underline{\mu_2}\mu_3 \mu_4 \mu_5}
}
\nonumber\\
&&
+{\cal F}_{ab}(x)^{\mu_3}R^{(ax,\,bx)\underline{\mu_1\mu_2}\mu_4 \mu_5}
+{\cal F}_{ab}(x)^{\mu_3\mu_4}R^{(ax,\,bx)\underline{\mu_1\mu_2}\mu_5} )
\nonumber\\
&&
+g_{\mu_1 \mu_2}g_{\mu_3 \mu_5}g_{\mu_4 \mu_6}
({\cal F}_{ab}(x)^{\mu_1}R^{(ax,\,bx)\underline{\mu_2}\mu_3 \mu_4 \mu_5\mu_6}
+{\cal F}_{ab}(x)^{\mu_3}R^{(ax,\,bx)\underline{\mu_1\mu_2}\mu_4 \mu_5\mu_6}
\nonumber\\
&&
+{\cal F}_{ab}(x)^{\mu_3\mu_4}R^{(ax,\,bx)\underline{\mu_1\mu_2}\mu_5\mu_6})\,]
\label{Hproducto}
\end{eqnarray}
Next we have to evaluate the action of ${\cal F}_{ab}(x)$ onto the
two and three point propagators. The following results are obtained

\begin{eqnarray}
\lefteqn{ {\cal F}_{ab}{^{\mu_1}}(x) \, g_{\mu_1 \mu_2} =
-\epsilon_{ab a_{2}} \delta(x-x_2)
- \partial{_{a_{2}}} g_{ax \, bx_{2}} \label{F1g} }\\
\lefteqn{ {\cal F}_{ab}{^{\mu_1}}(x) \, h_{\mu_1 \mu_2 \mu_3 } =
- g_{\mu_2  [\,ax} g_{\, bx] \, \mu_3} + (g_{ax \, bx_{2}} -
g_{ax \, bx_{3}}) g_{\mu_2 \mu_3} }  \nonumber\\
&&
+ {\textstyle{1 \over 2}} g_{ax  \,  bz}  \epsilon^{c_1 c_2 c_3}
[g_{\mu_3 \,c_1 z}\partial_{a_{2}}  \, g_{c_2 x_{2} \,c_3 z} -
g_{\mu_2 \,c_1 z}\partial_{a_{3}}  \, g_{c_2 x_{3} \,c_3 z} ]
\label{F1h}\\
\lefteqn{  {\cal F}_{ab}{^{\mu_1 \mu_2}}(x) \, g_{ \mu_1 \mu_3 }
g_{\mu_2 \mu_4}  =   g_{\mu_3  [ ax } \, g_{\, bx] \, \mu_4 } }
\label{F2g}\\
\lefteqn{  {\cal F}_{ab}{^{\mu_1 \mu_2}}(x) \, h_{ \mu_1 \mu_2 \mu_3}
 =   2\, h_{ax \,  bx \, \mu_3} } \label{F2h}
\end{eqnarray}
In the last term of (\ref{F1h}) an integration in $z$ is assumed.
Introducing (\ref{F1g})-(\ref{F2h}) into (\ref{Hproducto}) and
performing the integrations by parts indicated by the derivatives we get

\begin{eqnarray}
\lefteqn{ \textstyle{1\over9}H_0\,(\varphi_G * J_2) ({\bf R}) =
\epsilon_{abc}h_{\mu_1 \mu_2 \mu_3}
R^{(ax,\,bx)\underline{cx}\,\mu_1 \mu_2 \mu_3} }
\nonumber\\
&&+\epsilon_{abc}g_{\mu_1 \mu_3}g_{\mu_2 \mu_4}
[R^{(ax,\,bx)\underline{cx}\,\mu_1 \mu_2 \mu_3\mu_4}+
R^{(ax,\,bx)\underline{\mu_1\mu_3}\mu_2 \,cx\,\mu_4}]\nonumber\\
&&-g_{\mu_1 \mu_2}[2 h_{ax\,bx\,\mu_3}-
\epsilon^{c_1 c_2 c_3}g_{ax\,bz}
g_{\mu_3 \,c_1 z} g_{c_2 x \,c_3 z}]
R^{(ax,\,bx)\underline{\mu_1\mu_2}\mu_3}\nonumber\\
&&+g_{\mu_1 \mu_2}g_{\mu_3 \mu_4}
[g_{ax\,bz}\partial_{cz}
R^{(ax,\,bx)\underline{\mu_1\mu_2}\mu_3 \,cz\,\mu_4}\!-\!
(g_{ax\,bx_3}\!-\!g_{ax\,bx_4})
R^{(ax,\,bx)\underline{\mu_1\mu_2}\mu_3 \mu_4}]\nonumber\\
&&+g_{ax\,bz}[g_{\mu_1 \mu_3}g_{\mu_2 \mu_4}
\partial_{cz}R^{(ax,\,bx)\underline{cz}\,\mu_1 \mu_2 \mu_3\mu_4}+
h_{\mu_1 \mu_2 \mu_3}\partial_{cz}
R^{(ax,\,bx)\underline{cz}\,\mu_1 \mu_2 \mu_3}]
\end{eqnarray}
It is easy to demonstrate that
\begin{eqnarray}
2h_{ax\,bx\,\mu_3}&=&\epsilon^{c_1 c_2 c_3}g_{ax\,bz}
g_{\mu_3 \,c_1 z} g_{c_2 x \,c_3 z}\;,\label{haxbx}\\
\partial_{cz}R^{(ax,\,bx)\underline{\mu_1\mu_2}\mu_3 \,cz\,\mu_4}&=&
[\delta(z-x_3)-\delta(z-x_4)]
R^{(ax,\,bx)\underline{\mu_1\mu_2}\mu_3 \mu_4}\;,
\end{eqnarray}
and
\begin{equation}
\partial_{cz}R^{(ax,\,bx)\underline{cz}\,\mu_1 \ldots \mu_n}=0\;.
\end{equation}
Using these results one obtains
\begin{eqnarray}
\lefteqn{ \textstyle{1\over9}H_0\,(\varphi_G * J_2) ({\bf R}) =
\epsilon_{abc}h_{\mu_1 \mu_2 \mu_3}
R^{(ax,\,bx)\underline{cx}\,\mu_1 \mu_2 \mu_3} }
\nonumber\\
&&+\epsilon_{abc}g_{\mu_1 \mu_3}g_{\mu_2 \mu_4}
[R^{(ax,\,bx)\underline{cx}\,\mu_1 \mu_2 \mu_3\mu_4}+
R^{(ax,\,bx)\underline{\mu_1\mu_3}\mu_2 \,cx\,\mu_4}]
\label{HGJ2}
\end{eqnarray}
The determinant of the three metric on $J_2$ gives
\begin{equation}
det \,q \, J_2({\bf R}) =-6
\epsilon_{abc}[h_{\mu_1 \mu_2 \mu_3} R^{(ax,\, bx, \,cx)\mu_1 \mu_2 \mu_3}+
g_{\mu_1 \mu_3}g_{\mu_2 \mu_4} R^{(ax,\, bx, \,cx)\mu_1 \mu_2 \mu_3\mu_4}]
\label{detqJ2}
\end{equation}
In the appendix I it is demonstrated that the three-point-R that appears
in $det\,q$ can be related to the two-point-R of the Hamiltonian in the
following way

\begin{equation}
\epsilon_{abc}R^{ (ax,\, bx, \,cx) \b \mu} =
-\epsilon_{abc}[R^{(ax,\,bx)
\underline{\underline{cx}}\b \mu}+
\textstyle{1\over2}R^{(ax,\,bx)\underline{cx}\b \mu}]
\label{ultima}
\end{equation}
if the set of indices $\b \mu$ is cyclic and
\begin{equation}
R^{(ax,\,bx)\underline{cx}\b \mu}:=
\DG_{\b \alpha \b \beta}^{\b \mu}\,R^{(ax,\,bx)\b \alpha\,cx\,\b \beta}
\label{rabc1}
\end{equation}
\begin{equation}
R^{(ax,\,bx)\underline{\underline{cx}}\b \mu}:=
\DG_{\b \alpha \bb \beta}^{\b \mu}\,R^{(ax,\,bx)\b \alpha\,cx\,\b \beta}
\label{rabc2}
\end{equation}
Notice that the above expressions correspond to the algebraic constraint
combination (\ref{ac}) underlying the index $cx$ (in (\ref{rabc2}) one
has besides to overline the set of indices that follows $cx$). Using
(\ref{ultima}) we get from (\ref{HGJ2}) and (\ref{detqJ2})

\begin{eqnarray}
\lefteqn{ H_0\,(\varphi_G * J_2) -det \,q (x)\, J_2=9\epsilon_{abc}
g_{\mu_1 \mu_2}g_{\mu_3 \mu_4}
R^{(ax,\,bx)\underline{\mu_1\mu_2}\mu_3 \,cx\,\mu_4} }
\nonumber\\
&&+6\epsilon_{abc}h_{\mu_1 \mu_2 \mu_3}
[R^{(ax,\,bx)\underline{cx}\,\mu_1 \mu_2 \mu_3}-
R^{(ax,\,bx)\underline{\underline{cx}}\,\mu_1 \mu_2 \mu_3}]
\nonumber\\
&&+6\epsilon_{abc}g_{\mu_1 \mu_3}g_{\mu_2 \mu_4}
[R^{(ax,\,bx)\underline{cx}\,\mu_1 \mu_2 \mu_3\mu_4}-
R^{(ax,\,bx)\underline{\underline{cx}}\,\mu_1 \mu_2 \mu_3\mu_4}]
\label{HGJ2-detqJ2}
\end{eqnarray}
It is easy to see that the first contribution of the r.h.s. can be put
in the form

\begin{eqnarray}
&&\epsilon_{abc}g_{\mu_1 \mu_2}g_{\mu_3 \mu_4}
R^{(ax,\,bx)\underline{\mu_1\mu_2}\mu_3 \,cx\,\mu_4}=
2\epsilon_{abc}g_{\mu_1 \mu_2}g_{\mu_3 \mu_4}
[R^{(ax,\,bx)\mu_1\,cx\,(\mu_2 \mu_3\mu_4)_c}
\nonumber\\
&&\hspace{1.313cm}+R^{(ax,\,bx)\mu_1\mu_3\,cx\,\mu_2\mu_4}+
R^{(ax,\,bx)\mu_1\mu_3\,cx\,\mu_4\mu_2}]
\end{eqnarray}
and that
\begin{equation}
\epsilon_{abc}g_{\mu_1 \mu_2}g_{\mu_3 \mu_4}
R^{(ax,\,bx)\mu_1\,cx\,(\mu_2 \mu_3\mu_4)_c}\equiv0
\end{equation}
by symmetry considerations. These facts allow to write the
following expression for the Hamiltonian on the third coefficient of the
Jones polynomial

\begin{eqnarray}
\lefteqn{ H_0\,J_3 ({\bf R})=2\epsilon_{abc}
\{\,J_2 [\,{\bf R}^{(ax,\,bx)\underline{\underline{cx}}}\,] -
J_2 [\,{\bf R}^{(ax,\,bx)\underline{cx}}\,]  \} }
\nonumber\\
&&+18\epsilon_{abc}g_{\mu_1\mu_2}g_{\mu_3\mu_4}
[R^{(ax,\,bx)\mu_1 \mu_3 \,cx\, \mu_2 \mu_4}
+R^{(ax,\,bx)\mu_1  \mu_3 \,cx\, \mu_4 \mu_2}]
\label{newresult}
\end{eqnarray}
How can this result be zero? The only general way to get the
cancellation of (\ref{newresult}) is by using symmetry considerations.
The answer is negative: the Hamiltonian does not annihilates the third
coefficient of the Jones polynomial for general extended loops. In fact,
developing the two-point-R's according to (\ref{dos}) and using the
symmetry properties of the $R$, the propagators and $\epsilon_{abc}$ one
gets \cite{Gr}

\begin{eqnarray}
\lefteqn{ \textstyle{1\over24}H_0\,J_3 ({\bf R})=
\epsilon_{abc}h_{\mu_1\mu_2\mu_3}\,
R^{(ax\,\mu_1 \,bx\, \mu_2 \,cx\,\mu_3)_c} }
\nonumber\\
&&-3\epsilon_{abc}g_{\mu_1\mu_2}g_{\mu_3\mu_4}
\,[ R^{(ax\,\mu_1 \mu_3 \,bx\, \mu_2 \mu_4\,cx)_c}
+R^{(ax\,\mu_1 \mu_3 \,bx\, \mu_4 \mu_2\,cx)_c}
\nonumber\\
&&-R^{(ax\,\mu_1\,bx\,\mu_3\mu_2\,cx\,\mu_4)_c}]
\label{finalresult}
\end{eqnarray}
No further reduction is possible. In the next section we shall see that
new possibilities appear when (\ref{newresult}) is specialized to
ordinary loops.

\section{$H_0\,J_3$ in terms of ordinary loops}

The relationship between the extended loop and the loop representations
can be formulated in general. It is demonstrated that the constraints
(\ref{extdif}) and (\ref{ham}) generate those of the conventional loop
respresentation when extended loops are reduced to ordinary loops
\cite{DiGaGr2}. We will use here the same procedure of reduction to
express (\ref{newresult}) in terms of ordinary loops (that is, when one
imposes that ${\bf R} = {\bf R}(\gamma)$ is a multitangent field).

Let us consider a loop $\gamma$ that intersects itself $p$ times at the
spatial point $x$ (we say that the loop has ``multiplicity" $p$ at $x$).
Following \cite{DiGaGr2} we write

\begin{equation}
\gamma_o = \gamma^{(1)}{_o^x}  [\gamma_{xx}]_2^p
\gamma^{(1)}{_x^o}
\end{equation}
where $o$ is the origin of the loop and $\gamma_y^z$ indicates an open
path from $y$ to $z$. The loops $\gamma^{(i)}_{xx},\,i=1,\ldots,p$ are
the $p$ ``petals" (basepointed at $x$) of $\gamma_0$ and

\begin{equation}
[\gamma_{xx}]_i^{k} := \gamma^{(i)}_{xx}\gamma^{(i+1)}_{xx}
\cdots \gamma^{(k)}_{xx}
\end{equation}
Usually we denote this composition of loops simply by $\gamma_{ik}$. The
loop $\gamma_o$ is completely described by the multitangent fields
$X^{\b \mu}(\gamma_o)$ of all rank. As we know, these fields satisfy
the algebraic and differential constraints. Besides, the multitangents
have another property related to the possibility to write a loop as
a composition of open paths. In general

\begin{equation}
X^{\mu_1 \ldots \mu_n}(\gamma_o) = \int_{\gamma_o} dz^{a_i} \delta(x_i -
z) X^{\mu_1 \ldots \mu_{i-1}}(\gamma_o^z)\,X^{\mu_{i+1}
\ldots \mu_n}(\gamma_z^o)
\end{equation}
If the index $\mu_i$ is fixed at the point $x$ one can write

\begin{eqnarray}
\lefteqn{
X^{\mu_1 \ldots \mu_i \, ax \,\mu_{i+1} \ldots \mu_n}(\gamma_o) = }
\nonumber \\
&& \sum_{m=1}^{p}
X^{\mu_1 \ldots \mu_i}(\gamma^{(1)}{_o^x} [\gamma_{xx}]_2^m) \,
T_m^{ax} \,
X^{\mu_{i+1} \ldots \mu_n}([\gamma_{xx}]_{m+1}^p \gamma^{(1)}{_x^o})
\label{tangent}
\end{eqnarray}
where $T_m^{ax}$ is the tangent at $x$ when the loop crosses the time
$m$ to this point (in the above expression the following convention is
assumed: $[\gamma_{xx}]_{m+1}^m \approx {\em i}_{xx}$, with ${\em
i}_{xx}$ the null path). The property (\ref{tangent}) can be easily
generalized to the case of any number of spatial indices evaluated at
$x$. In the appendix II we shall show that the following decomposition
is valid for $R^{(ax,\,bx)\b \alpha\,cx\, \b \beta}(\gamma_o)$:

\begin{eqnarray}
\lefteqn{ \epsilon_{abc}
R^{(ax,\,bx)\b \alpha\,cx\,\b \beta}(\gamma_o)=-\textstyle{1\over2}
\epsilon_{abc}\sum_{m,q,r}T_m^{ax}T_q^{bx}T_r^{cx}\times }
\nonumber\\
&&[X^{(\b \alpha\mid}(\gamma_{mq}\overline{\gamma}_{rm})
X^{\mid\bb \beta)_c}(\gamma_{qr})+
X^{(\b \alpha\mid}(\overline{\gamma}_{rm}\gamma_{mq})
X^{\mid\bb \beta)_c}(\overline{\gamma}_{qr})+
\nonumber\\
&&X^{(\b \alpha\mid}(\gamma_{qr}\overline{\gamma}_{mq})
X^{\mid\bb \beta)_c}(\gamma_{rm})+
X^{(\b \alpha\mid}(\overline{\gamma}_{mq}\gamma_{qr})
X^{\mid\bb \beta)_c}(\overline{\gamma}_{rm})+
\nonumber\\
&&X^{(\b \alpha\mid}(\gamma_{rm}\overline{\gamma}_{qr})
X^{\mid\bb \beta)_c}(\gamma_{mq})+
X^{(\b \alpha\mid}(\overline{\gamma}_{qr}\gamma_{rm})
X^{\mid\bb \beta)_c}(\overline{\gamma}_{mq})]
\label{rabc3}
\end{eqnarray}
where $(\cdot\mid\mid\cdot)_c$ indicates the cyclic permutation {\it of
the sets} of indices and the sums run from $1$ to $p-2$ for $m$, from
$m+1$ to $p-1$ for $q$ and from $q+1$ to $p$ for $r$. In the above
expression $\overline{\gamma}_{ik}:= \overline{\gamma}^{(k)}_{xx}
\overline{\gamma}^{(k-1)}_{xx} \cdots \overline{\gamma}^{(i)}_{xx}$
defines the rerouting of the loop $\gamma_{ik}$ and

\begin{eqnarray}
\gamma_{mq} &:=& [\gamma_{xx}]_{m+1}^q\\
\gamma_{qr} &:=& [\gamma_{xx}]_{q+1}^r\\
\gamma_{rm} &:=& [\gamma_{xx}]_{r+1}^p
\gamma^{(1)}{_x^o}\gamma^{(1)}{_o^x}  [\gamma_{xx}]_2^m
\end{eqnarray}
Notice now in which way the combinations defined by (\ref{rabc1}) and
(\ref{rabc2}) are generated from (\ref{rabc3}): one has simply to
contract (\ref{rabc3}) with $\DG_{\b \alpha \b \beta}^{\b \mu}$ and
$\DG_{\b \alpha \bb \beta}^{\b \mu}$ respectively. This means that a
group product is formed for all the terms of (\ref{rabc3}) and the only
difference between the two cases will be the ``rerouting" of the set of
indices $\b \beta$. Using the fact that

\begin{equation}
\DG_{\b \alpha \b \beta}^{\b \mu}\,X^{\b \alpha}(\gamma)\,
X^{\b \beta}(\gamma')=X^{\b \mu}(\gamma\gamma')\;\,,
\end{equation}
and
\begin{equation}
X^{\bb \mu}(\gamma)=X^{\b \mu}(\overline{\gamma})\;,
\label{reruteo}
\end{equation}
one gets from (\ref{rabc3}) the following results:
\begin{eqnarray}
\lefteqn{ \epsilon_{abc}
R^{(ax,\,bx)\underline{cx}\, \b \mu}(\gamma_o)=
-\epsilon_{abc}\sum_{m,q,r}
T_m^{ax}T_q^{bx}T_r^{cx}\times }
\nonumber\\
&&[R^{\b \mu}(\overline{\gamma}_{mq}\gamma_{qr}\gamma_{rm})+
R^{\b \mu}(\overline{\gamma}_{qr}\gamma_{rm}\gamma_{mq})+
R^{\b \mu}(\overline{\gamma}_{rm}\gamma_{mq}\gamma_{qr})+
\nonumber\\
&&R^{\b \mu}(\gamma_{mq}\gamma_{qr}\overline{\gamma}_{rm})+
R^{\b \mu}(\gamma_{qr}\gamma_{rm}\overline{\gamma}_{mq})+
R^{\b \mu}(\gamma_{rm}\gamma_{mq}\overline{\gamma}_{qr})]
\label{rloop1}
\end{eqnarray}
and
\begin{eqnarray}
\lefteqn{
\epsilon_{abc}R^{(ax,\,bx)\underline{\underline{cx}}\,\b \mu}(\gamma_o)=
-\epsilon_{abc}\sum_{m,q,r}T_m^{ax}T_q^{bx}T_r^{cx}\times }
\nonumber\\
&&\left[R^{\b \mu}(\gamma_{mq}\overline{\gamma}_{rm}\gamma_{qr})+
R^{\b \mu}(\gamma_{qr}\overline{\gamma}_{mq}\gamma_{rm})+
R^{\b \mu}(\gamma_{rm}\overline{\gamma}_{qr}\gamma_{mq})\right.
\nonumber\\
&&
+\textstyle{1\over2}\{X^{\b \mu}(\gamma_{mq}\gamma_{rm}\overline{\gamma}_{qr})
+X^{\b \mu}(\overline{\gamma}_{qr}\gamma_{mq}\gamma_{rm})\}
\nonumber\\
&&
+\textstyle{1\over2}\{X^{\b \mu}(\gamma_{qr}\gamma_{mq}\overline{\gamma}_{rm})
+X^{\b \mu}(\overline{\gamma}_{rm}\gamma_{qr}\gamma_{mq})\}
\nonumber\\
&&\left.
+\textstyle{1\over2}\{X^{\b \mu}(\gamma_{rm}\gamma_{qr}\overline{\gamma}_{mq})
+X^{\b \mu}(\overline{\gamma}_{mq}\gamma_{rm}\gamma_{qr})\}\right]
\label{rloop2}
\end{eqnarray}
We see that the loop $\gamma_o$ is descomposed into a ``three petal
structure" with a rerouted portion. Each ``petal" consists of a
combination of loops basepointed at $x$. Suppose now that the greek
indices of the above expressions are contracted with suitable
propagators $D_{\b \mu}$. Then using the cyclicity of $D_{\b \mu}$ we
get

\begin{eqnarray}
\lefteqn{ \epsilon_{abc}D_{\b \mu}
R^{(ax,\,bx)\underline{cx}\, \b \mu}(\gamma_o)=
-2\epsilon_{abc}\sum_{m,q,r}
T_m^{ax}T_q^{bx}T_r^{cx}\times }
\nonumber\\
&&[\psi(\overline{\gamma}_{mq}\gamma_{qr}\gamma_{rm})+
\psi(\gamma_{mq}\overline{\gamma}_{qr}\gamma_{rm})+
\psi(\gamma_{mq}\gamma_{qr}\overline{\gamma}_{rm})]
\label{rloop3}
\end{eqnarray}
and
\begin{eqnarray}
\lefteqn{ \epsilon_{abc}D_{\b \mu}
R^{(ax,\,bx)\underline{\underline{cx}}\,\b \mu}(\gamma_o)=
-2\epsilon_{abc}\sum_{m,q,r}T_m^{ax}T_q^{bx}T_r^{cx}\times }
\nonumber\\
&&[\psi(\overline{\gamma}_{mq}\gamma_{rm}\gamma_{qr})+
\psi(\gamma_{rm}\overline{\gamma}_{qr}\gamma_{mq})+
\psi(\gamma_{qr}\gamma_{mq}\overline{\gamma}_{rm})]
\label{rloop4}
\end{eqnarray}
with
\begin{equation}
\psi(\gamma):=D_{\b \mu}R^{\b \mu}(\gamma)
\end{equation}
If $\psi(\gamma)$ is a knot invariant, the r.h.s of equations
(\ref{rloop3}) and (\ref{rloop4}) reduces to a combination of invariants
evaluated onto the three petal structure. This result shows that the
algebraic combinations (\ref{rabc1}) and (\ref{rabc2}) are able to
capture relevant geometrical information when they are specialized to
ordinary loops.

What happens with the terms of the form
$g_{\cdot\cdot}g_{\cdot\cdot}R^{(ax,\,bx)\cdot\cdot\,cx\,\cdot\cdot}$ in
(\ref{newresult})? These terms generate Gauss link invariants when
ordinary loops are introduced. In effect, it is possible to show that

\begin{eqnarray}
\lefteqn{ \epsilon_{abc}g_{\mu_1\mu_2}g_{\mu_3\mu_4}
[R^{(ax,\,bx)\mu_1 \mu_3 \,cx\, \mu_2 \mu_4}+
R^{(ax,\,bx)\mu_1 \mu_3 \,cx\, \mu_4 \mu_2}]= }
\nonumber\\
&&-2\epsilon_{abc}\sum_{m,q,r}
T_m^{ax}T_q^{bx}T_r^{cx}g_{\mu_1\mu_2}g_{\mu_3\mu_4}\times
\nonumber\\
&&\left[R^{\mu_1 \mu_3}(\gamma_{mq})
\{R^{\mu_2 \mu_4}(\overline{\gamma}_{qr}\gamma_{rm})+
R^{\mu_2 \mu_4}(\gamma_{rm}\overline{\gamma}_{qr})\}\right.
\nonumber\\
&&+R^{\mu_1 \mu_3}(\gamma_{qr})
\{R^{\mu_2 \mu_4}(\overline{\gamma}_{rm}\gamma_{mq})+
R^{\mu_2 \mu_4}(\gamma_{mq}\overline{\gamma}_{rm})\}
\nonumber\\
&&+R^{\mu_1 \mu_3}(\gamma_{rm})
\{R^{\mu_2 \mu_4}(\overline{\gamma}_{mq}\gamma_{qr})+
R^{\mu_2 \mu_4}(\gamma_{qr}\overline{\gamma}_{mq})\}\left.\right]
\label{link1}
\end{eqnarray}
Due to the algebraic constraint,
\begin{equation}
R^{\mu_1 \mu_3}(\gamma)=
\textstyle{1\over2}X^{\mu_1}(\gamma)X^{\mu_3}(\gamma)
\end{equation}
and
\begin{equation}
R^{\mu_2 \mu_4}(\gamma\overline{\gamma}')= \textstyle{1\over2}
[X^{\mu_2}(\gamma)-X^{\mu_2}(\gamma')]
[X^{\mu_4}(\gamma)-X^{\mu_4}(\gamma')]
\end{equation}
So one can write

\begin{equation}
g_{\mu_1\mu_2}g_{\mu_3\mu_4}R^{\mu_1 \mu_3}(\gamma)
R^{\mu_2 \mu_4}(\gamma'\overline{\gamma}'')=
\textstyle{1\over3}[\varphi_G(\gamma,\gamma') -
\varphi_G(\gamma,\gamma'')]^2
\label{link2}
\end{equation}
with

\begin{equation}
\varphi_G(\gamma,\gamma'):=\textstyle{3\over4}
g_{\alpha_1 \alpha_2}X^{\alpha_1}(\gamma)X^{\alpha_2}(\gamma')
\end{equation}
the Gauss linking number of the loops $\gamma$ and $\gamma'$.
Introducing now equations (\ref{rloop3}), (\ref{rloop4}), (\ref{link1})
and (\ref{link2}) into (\ref{newresult}) the following result is
obtained

\begin{eqnarray}
\lefteqn{ H_0\,J_3 (\gamma_o)=4\epsilon_{abc}\sum_{m,q,r}
T_m^{ax}T_q^{bx}T_r^{cx}\times }
\nonumber\\
&&\{
J_2(\overline{\gamma}_{mq}\gamma_{qr}\gamma_{rm})+
J_2(\gamma_{mq}\overline{\gamma}_{qr}\gamma_{rm})+
J_2(\gamma_{mq}\gamma_{qr}\overline{\gamma}_{rm})
\nonumber\\
&&-J_2(\overline{\gamma}_{mq}\gamma_{rm}\gamma_{qr})
-J_2(\gamma_{rm}\overline{\gamma}_{qr}\gamma_{mq})
-J_2(\gamma_{qr}\gamma_{mq}\overline{\gamma}_{rm})
\nonumber\\
&&-6[\varphi_G(\gamma_{mq},\gamma_{qr})-\varphi_G(\gamma_{mq},\gamma_{rm})]^2
-6[\varphi_G(\gamma_{qr},\gamma_{rm})-\varphi_G(\gamma_{qr},\gamma_{mq})]^2
\nonumber\\
&&-6[\varphi_G(\gamma_{rm},\gamma_{mq})-\varphi_G(\gamma_{rm},\gamma_{qr})]^2
\left.\right\}
\end{eqnarray}
This expression can be simplified using the following
Mandelstam identity valid for $J_2$:

\begin{equation}
J_2(\gamma_{mq}\gamma_{qr}\overline{\gamma}_{rm})-
J_2(\gamma_{qr}\gamma_{mq}\overline{\gamma}_{rm})=
J_2(\gamma_{qr}\gamma_{mq}\gamma_{rm})-
J_2(\gamma_{mq}\gamma_{qr}\gamma_{rm})
\label{manident}
\end{equation}
We then conclude

\begin{eqnarray}
\lefteqn{ H_0\,J_3 (\gamma_o)=-12\epsilon_{abc}\sum_{m,q,r}
T_m^{ax}T_q^{bx}T_r^{cx}\left\{\right.
J_2(\gamma_{mq}\gamma_{qr}\gamma_{rm})-
J_2(\gamma_{qr}\gamma_{mq}\gamma_{rm}) }
\nonumber\\
&&\hspace{0.2cm}
+2[\varphi_G(\gamma_{mq},\gamma_{qr})-\varphi_G(\gamma_{mq},\gamma_{rm})]^2
+2[\varphi_G(\gamma_{qr},\gamma_{rm})-\varphi_G(\gamma_{qr},\gamma_{mq})]^2
\nonumber\\
&&\hspace{0.2cm}
+2[\varphi_G(\gamma_{rm},\gamma_{mq})-\varphi_G(\gamma_{rm},\gamma_{qr})]^2
\left.\right\}
\label{hj3loops}
\end{eqnarray}
This expression has a quite nontrivial geometrical content. The action
of the Hamiltonian on the third coefficient of the Jones polynomial is
given by a combination of knot $J_2$ and link $\varphi_G$ invariants for
a loop with an intersection of arbitrary multiplicity. Moreover, the
knot and link invariants are evaluated onto a precise decomposition of
the original loop into a three petal structure basepointed at $x$.

The knot and link invariants appear combined into pairs. This fact
suggests that $J_3$ could in principle be annihilated by $H_0$ by means
of simple topological requirements. For example, for the unknot trefoil
(\ref{hj3loops}) reduces to

\begin{equation}
H_0\,J_3({\mbox{\scriptsize{unknot\,trefoil}}})=
12\epsilon_{abc}T_1^{ax}T_2^{bx}T_3^{cx}
\{J_2(\gamma_{2}\gamma_{1}\gamma_{3})-
J_2(\gamma_{1}\gamma_{2}\gamma_{3})\}
\label{trefoil}
\end{equation}
that is nonzero in general. In spite that the cancellation does not take
place for the simplest three petal structure, it seems plausible that it
could happen for some topologies of that kind. A first approach to the
problem has not revealed any immediate solution of this type. This topic
is currently under progress.

\subsection{The Mandelstam identities of $J_3$}
The Mandesltam identites \cite{mand} are requisites for the
wavefunctions in the loop and extended loop representations (the
identities follows from the properties of the Wilson loop and the Wilson
functional is in the basis of the loop and extended loop transforms).
One can see that $J_3$ does not satisfies all the Mandelstam identities
in general (that is, for arbitrary loops or extended loops) \footnote{I
thank Rodolfo Gambini to point me out this fact.}. From this point of
view one can question the expectative that this knot invariant could
represent a genuine quantum state of gravity \footnote{This fact was not
realized at the time the conjecture mentioned in Sect. 2 was proposed.}.

An intriguing fact that we are going to consider here is that the
Mandelstam identities are recovered totally by the $J_3$ invariant {\it
if} the topological conditions necessary for $H_0\,J_3=0$ are fulfilled.

As it was mentioned in Sect. 2, the Kauffman bracket can be viewed as
the expectation value of the Wilson loop. This means that the knot
invariants $K_m(\gamma)$ of the expansion (\ref{hkauff1}) satisfy the
Mandelstam identities by construction. According to (\ref{km}), each
coefficient of the expansion is expressed by a sum of products of Gauss
and Jones invariants. As the Mandelstam identities are nonlinear, the
identities will not be inherited by $J_n(\gamma)$ in general
\footnote{The second coefficient $J_2$ is an exception to this rule.}
(the abelian property makes trivial the Mandelstam identities for the
case of the Gauss invariants).

Let us limit the discussion to the case of interest. To third order one
has

\begin{equation}
K_3 = J_3 - J_2\,\varphi_G - \textstyle{1\over{3!}}
\varphi_G^3
\end{equation}
The product $J_2\,\varphi_G$ is responsible that the property
(\ref{manident}) will not be inherited by $J_3$ \footnote{There are
three basic Mandelstam identities: the cyclicity
($J_3(\gamma_1\gamma_2)=J_3(\gamma_2\gamma_1)$), the inversion
($J_3(\gamma)=J_3(\overline{\gamma})$) and (\ref{manident}). The first
two are valid in general for $J_3$.}. However, from the fact that $J_2$
and $\varphi_G$ satisfy the identity (\ref{manident}) it is
straightforward to derive the following relationship for the product of
the two invariants

\begin{eqnarray}
\hspace{-0.5cm}
\lefteqn{ (J_2\,\varphi_G)(\gamma_{mq}\gamma_{qr}\gamma_{rm})= }
\nonumber\\
&&\hspace{-0.5cm}
(J_2\,\varphi_G)(\gamma_{qr}\gamma_{mq}\gamma_{rm})
+(J_2\,\varphi_G)(\gamma_{qr}\gamma_{mq}\overline{\gamma}_{rm})
-(J_2\,\varphi_G)(\gamma_{mq}\gamma_{qr}\overline{\gamma}_{rm})+
\nonumber\\
&&\hspace{-0.5cm}
[J_2(\gamma_{qr}\gamma_{mq}\overline{\gamma}_{rm})-
J_2(\gamma_{mq}\gamma_{qr}\overline{\gamma}_{rm})]
[\varphi_G(\gamma_{qr}\gamma_{mq}\gamma_{rm})-
\varphi_G(\gamma_{mq}\gamma_{qr}\overline{\gamma}_{rm})]
\label{mandpro2}
\end{eqnarray}
where we have used the fact that
\begin{equation}
\varphi_G(\gamma_{mq}\gamma_{qr}\gamma_{rm})
=\varphi_G(\gamma_{qr}\gamma_{mq}\gamma_{rm})
\end{equation}
The difference of the Gauss invariants is simply given by
\begin{equation}
\varphi_G(\gamma_{qr}\gamma_{mq}\gamma_{rm})
-\varphi_G(\gamma_{mq}\gamma_{qr}\overline{\gamma}_{rm})=
4[\varphi_G(\gamma_{mq},\gamma_{rm})
+\varphi_G(\gamma_{qr},\gamma_{rm})]
\end{equation}
and using (\ref{manident}) again we conclude from (\ref{mandpro2})
\begin{eqnarray}
\lefteqn{ (J_2\,\varphi_G)(\gamma_{mq}\gamma_{qr}\gamma_{rm})+
(J_2\,\varphi_G)(\gamma_{mq}\gamma_{qr}\overline{\gamma}_{rm})= }
\nonumber\\
&&(J_2\,\varphi_G)(\gamma_{qr}\gamma_{mq}\gamma_{rm})+
(J_2\,\varphi_G)(\gamma_{qr}\gamma_{mq}\overline{\gamma}_{rm})+
\nonumber\\
&&4[J_2(\gamma_{mq}\gamma_{qr}\gamma_{rm})-
J_2(\gamma_{qr}\gamma_{mq}\gamma_{rm})]
[\varphi_G(\gamma_{mq},\gamma_{rm})+
\varphi_G(\gamma_{qr},\gamma_{rm})]
\label{mandproducto}
\end{eqnarray}
Those three petal structures that are annihilated by the Hamiltonian
constraint would verify that
\begin{equation}
J_2(\gamma_{mq}\gamma_{qr}\gamma_{rm})-
J_2(\gamma_{qr}\gamma_{mq}\gamma_{rm})=0
\end{equation}
For these cases, $J_2\varphi_G$ recovers the property
(\ref{manident}). As $\varphi_G^3$ satisfies the Mandels- tam
identities in general, we conclude that $J_3$ will verify the Mandelstam
identity (\ref{manident}) for those loops that makes
$H_0\,J_3(\gamma_o)=0$ at the intersecting points.

\section{Conclusions}
The initial question about the third coefficient of the Jones polynomial
in quantum gravity can not be answered with a simple yes or no. In the
analysis, we have passed from a negative answer in the extended loop
manifold to a new expectative in the ordinary loop space. The new
expectative is based in the nontrivial topological content of the result
(\ref{hj3loops}).

The possibility that this result could provide a new solution of the
Wheeler-DeWitt equation is still unclear. Besides the proper difficulty
associated with the topological conditions that have to be fulfilled,
there exists some problems on the general ground. The main problem
follows from the restriction of the domain of definition of the loop
wavefunction. The limitation of the domain in the loop space implies
some kind of characteristic function that takes the value one for a set
of loops with a definite topology and zero otherwise. This fact faces up
two new difficulties: the action of the Hamiltonian on the Heaviside
part of the wavefunction could be nontrivial and the Mandelstam
identities of the restricted wavefunction break down. Up to present it
is not clear how to solve these questions in a general way. It is worth
of emphasize that this objections are shared by the smoothened loops of
reference \cite{AsRoSm}.

To finalize, a comment about the use of extended loops in quantum
gravity is in order. In a companion article \cite{Gr} it is explicitely
proved that the Kauffman bracket is a solution of the Hamiltonian
constraint with cosmological constant to third order. This requires the
explicit computation of the vacuum Hamiltonian on $J_3$, a very involved
task from the point of view of the conventional loop representation
\footnote{In general, the loop derivative obligates to limit the
analysis to loops with some kind of simple intersection (typically,
intersections of multiplicity three). Notice that the results obtained
via extended loops are valid for arbitrary ordinary loops.}. Besides of
this, the analysis developed in \cite{Gr} shows that a systematic of
operation exists for the constraints in the extended loop framework.
This systematic allows to raise several interesting questions about knot
theory and quantum gravity, such as: Which are the analytical
expressions of the knot invariants in terms of the two and three point
propagators of the Chern-Simons theory that satisfy the Mandelstam
identities?; or: There exist loop wavefunctions of this type {\it
besides} the Kauffman bracket, the exponential of the Gauss number and
the second coefficient of the Alexander-Conway coefficient? These
questions are of interest to the knowledge of the state of space of
quantum gravity. Extended loops are able to provide an answer to these
(and related) topics.

\section*{Acknowledgments}
I want to thank Cayetano Di Bartolo and Rodolfo Gambini for many
fruitful discussions. Also, I specially thank Jorge Pullin for a
critical reading of the manuscript.

\section*{Appendix I}
The action of the determinant of the three metric is characterized in
the extended loop representation by ${\bf R}^{(ax,\,bx,\,cx)}$. We show
here that this combination of multitensor fields can be written in terms
of the two-point-R that appears in the Hamiltonian. The definition of
the three-point-R is

\begin{equation}
R^{ (ax,\, bx, \,cx) \b \mu} =
\DG_{\b \sigma_1 \b \sigma_2 \b \sigma_3}^{\b \mu}\,
[2\,R^{(ax\, \b \sigma_2 \,bx\, \b \sigma_1 \,cx\,\bb \sigma_3)_c}+
R^{(ax\, \b \sigma_1 \,bx\, \b \sigma_2 \,cx\,\bb \sigma_3)_c}]
\label{aI0}
\end{equation}
{}From (\ref{dos}) we have
\begin{equation}
R^{(ax, \,bx)\b \alpha\,cx\,\b \beta} =
\DG_{\b \pi \b \theta}^{\b \alpha\,cx\,\b \beta}\,
R^{(ax\, \b \pi \,bx\,\bb \theta)_c}
\end{equation}
and it is easy to see that
\begin{equation}
\DG_{\b \pi \b \theta}^{\b \alpha\,cx\,\b \beta}=
\DG_{\b \pi}^{\b \alpha\,cx\,\b \sigma}
\DG_{\b \sigma \b \theta}^{\b \beta}+
\DG_{\b \pi \b \sigma}^{\b \alpha}
\DG_{\b \theta}^{\b \sigma\,cx\,\b \beta}
\label{aI5}
\end{equation}
Then
\begin{equation}
R^{(ax, \,bx)\b \alpha\,cx\,\b \beta} =
\DG_{\b \sigma \b \theta}^{\b \beta}
R^{(ax\, \b \alpha\,cx\,\b \sigma \,bx\,\bb \theta)_c}-
\DG_{\b \pi \b \sigma}^{\b \alpha}
R^{(ax\, \b \pi\,bx\,\bb \beta \,cx\,\bb \sigma)_c}
\end{equation}
The $R$'s are invariant under the overline operation:
$R^{\bb \mu}\equiv R^{\b \mu}$. Using this property to reverse the
sequence of indices of the last term of the r.h.s we get

\begin{equation}
R^{(ax, \,bx)\b \alpha\,cx\,\b \beta} =
\DG_{\b \sigma_1 \b \sigma_2}^{\b \beta}
R^{(ax\, \b \alpha\,cx\,\b \sigma_1 \,bx\,\bb \sigma_2)_c}+
\DG_{\b \sigma_1 \b \sigma_2}^{\b \alpha}
R^{(ax\, \b \sigma_2\,cx\,\b \beta \,bx\,\bb \sigma_1)_c}
\label{aI4}
\end{equation}
Then,
\begin{eqnarray}
R^{(ax,\,bx)\underline{cx}\b \mu}&:=&
\DG_{\b \alpha \b \beta}^{\b \mu}\,R^{(ax,\,bx)\b \alpha\,cx\,\b \beta}
\nonumber\\
&=&\DG_{\b \alpha \b \sigma_1 \b \sigma_2}^{\b \mu}
R^{(ax\, \b \alpha\,cx\,\b \sigma_1 \,bx\,\bb \sigma_2)_c}+
\DG_{\b \sigma_1 \b \sigma_2\b \beta}^{\b \mu}
R^{(ax\, \b \sigma_2\,cx\,\b \beta \,bx\,\bb \sigma_1)_c}
\label{aI1}
\end{eqnarray}
If the set of indices $\b \mu$ is cyclic one has
\begin{equation}
\DG_{\b \sigma_1 \b \sigma_2\b \beta}^{(\b \mu)_c}=
\DG_{\b \sigma_2 \b \beta\b \sigma_1}^{(\b \mu)_c}
\end{equation}
and then we get from (\ref{aI1})

\begin{equation}
\epsilon_{abc}R^{(ax,\,bx)\underline{cx}\b \mu}=
-2\epsilon_{abc} \DG_{\b \sigma_1 \b \sigma_2\b \sigma_3}^{\b \mu}
R^{(ax\, \b \sigma_1\,bx\,\b \sigma_2 \,cx\,\bb \sigma_3)_c}
\label{aI2}
\end{equation}
A similar procedure can be applied to (\ref{rabc2}). One obtains in this
case

\begin{equation}
\epsilon_{abc}R^{(ax,\,bx)\underline{\underline{cx}}\b \mu}
=-2\epsilon_{abc} \DG_{\b \sigma_1 \b \sigma_2\b \sigma_3}^{\b \mu}
R^{(ax\, \b \sigma_2\,bx\,\b \sigma_1 \,cx\,\bb \sigma_3)_c}
\label{aI3}
\end{equation}
Introducing (\ref{aI2}) and (\ref{aI3}) into (\ref{aI0}) we conclude
that
\begin{equation}
\epsilon_{abc}R^{ (ax,\, bx, \,cx) \b \mu} =-\epsilon_{abc}
[R^{(ax,\,bx)\underline{\underline{cx}}\b\mu}+
\textstyle{1\over2}R^{(ax,\,bx)\underline{cx}\b \mu}]
\end{equation}
if the set of indices $\b \mu$ is cyclic.

\section*{Appendix II}

In this appendix we shall demonstrate that
\begin{eqnarray}
\lefteqn{ \epsilon_{abc}
R^{(ax,\,bx)\b \alpha\,cx\,\b \beta}(\gamma_o)=-\textstyle{1\over2}
\epsilon_{abc}\sum_{m,q,r}T_m^{ax}T_q^{bx}T_r^{cx}\times }
\nonumber\\
&&[X^{(\b \alpha\mid}(\gamma_{mq}\overline{\gamma}_{rm})
X^{\mid\bb \beta)_c}(\gamma_{qr})+
X^{(\b \alpha\mid}(\overline{\gamma}_{rm}\gamma_{mq})
X^{\mid\bb \beta)_c}(\overline{\gamma}_{qr})+
\nonumber\\
&&X^{(\b \alpha\mid}(\gamma_{qr}\overline{\gamma}_{mq})
X^{\mid\bb \beta)_c}(\gamma_{rm})+
X^{(\b \alpha\mid}(\overline{\gamma}_{mq}\gamma_{qr})
X^{\mid\bb \beta)_c}(\overline{\gamma}_{rm})+
\nonumber\\
&&X^{(\b \alpha\mid}(\gamma_{rm}\overline{\gamma}_{qr})
X^{\mid\bb \beta)_c}(\gamma_{mq})+
X^{(\b \alpha\mid}(\overline{\gamma}_{qr}\gamma_{rm})
X^{\mid\bb \beta)_c}(\overline{\gamma}_{mq})]
\label{aII0}
\end{eqnarray}
for a loop with multiplicity $p$ at $x$. From (\ref{aI4}) we know that

\begin{equation}
\epsilon_{abc}R^{(ax, \,bx)\b \alpha\,cx\,\b
\beta}=-\epsilon_{abc} [\DG_{\b \sigma_1 \b \sigma_2}^{\b \alpha}
R^{(ax\, \b \sigma_2\,bx\,\b \beta \,cx\,\bb \sigma_1)_c}+ \DG_{\b
\sigma_1 \b \sigma_2}^{\b \beta} R^{(ax\, \b \alpha\,bx\,\b \sigma_1
\,cx\,\bb \sigma_2)_c}] \label{aII5}
\end{equation}
We start by considering in general the decomposition of $R^{(ax\, \b
\sigma_1\,bx\,\b \sigma_2 \,cx\,\b \sigma_3)_c}(\gamma_o)$. The
expression of this quantity in terms of multivector fields is

\begin{equation}
R^{(ax\, \b \sigma_1\,bx\,\b \sigma_2 \,cx\,\b
\sigma_3)_c}= \textstyle{1\over2}[ X^{(ax\, \b \sigma_1\,bx\,\b \sigma_2
\,cx\,\b \sigma_3)_c}- X^{(ax\, \bb \sigma_3\,cx\,\bb \sigma_2 \,bx\,\bb
\sigma_1)_c}] \label{aII4}
\end{equation}
Let us first develop $X^{(ax\, \b \sigma_1\,bx\,\b \sigma_2 \,cx\,\b
\sigma_3)_c}(\gamma_o)$. The cyclic combination of multivector fields
with one spatial index evaluated at $x$ can be written in this way

\begin{equation}
X^{(ax\,\b \sigma)_c}=\DG_{\b \lambda_1 \b \lambda_2}^{\b \sigma}
X^{\b \lambda_2 \,ax\,\b \lambda_1}
\end{equation}
Then

\begin{equation}
X^{(ax\, \b \sigma_1\,bx\,\b \sigma_2 \,cx\,\b \sigma_3)_c}
=\DG_{\b \lambda_1 \b \lambda_2}
^{\b \sigma_1\,bx\,\b \sigma_2 \,cx\,\b \sigma_3}
X^{\b \lambda_2 \,ax\,\b \lambda_1}
\label{aII1}
\end{equation}
The delta matrix with two spatial indices fixed at $x$ admits the
following decomposition

\begin{eqnarray}
\DG_{\b \lambda_1 \b \lambda_2}
^{\b \sigma_1\,bx\,\b \sigma_2 \,cx\,\b \sigma_3}&=&
\DG_{\b \lambda_1}^{\b \sigma_1\,bx\,\b \sigma_2 \,cx\,\b \rho_1}
\DG_{\b \rho_1 \b \lambda_2}^{\b \sigma_3}+
\DG_{\b \lambda_1}^{\b \sigma_1\,bx\,\b \rho_1}
\DG_{\b \rho_1 \b \rho_2}^{\b \sigma_2}
\DG_{\b \lambda_2}^{\b \rho_2\,cx\,\b \sigma_3}
\nonumber\\
&+&\DG_{\b \lambda_1 \b \rho_1}^{\b \sigma_1}
\DG_{\b \lambda_2}^{\b \rho_1\,bx\,\b \sigma_2\,cx\,\b \sigma_3}
\label{aII2}
\end{eqnarray}
Introducing (\ref{aII2}) into (\ref{aII1}) we get

\begin{eqnarray}
X^{(ax\, \b \sigma_1\,bx\,\b \sigma_2 \,cx\,\b \sigma_3)_c}
&=&\DG_{\b \lambda_1 \b \lambda_2}^{\b \sigma_1}
X^{\b \lambda_2 \,bx\,\b \sigma_2\,cx\,\b \sigma_3\,ax\,\b \lambda_1}+
\DG_{\b \lambda_1 \b \lambda_2}^{\b \sigma_2}
X^{\b \lambda_2 \,cx\,\b \sigma_3\,ax\,\b \sigma_1\,bx\,\b \lambda_1}
\nonumber\\
&+&\DG_{\b \lambda_1 \b \lambda_2}^{\b \sigma_3}
X^{\b \lambda_2 \,ax\,\b \sigma_1\,bx\,\b \sigma_2\,cx\,\b \lambda_1}
\label{aII3}
\end{eqnarray}
A multitangent field with three indices evaluated at $x$ decomposes
in the following way for a loop with multiplicity $p$ at $x$

\begin{eqnarray}
\lefteqn{
X^{\b \lambda_2 \,ax\,\b \sigma\,bx\,\b \sigma'\,cx\,\b \lambda_1}
(\gamma_o)= \sum_{m,q,r}\,T_m^{ax}\,T_q^{bx} \,T_r^{cx} \,\times }
\nonumber\\
&&X^{\b \lambda_2}(\gamma^{(1)}{_o^x} [\gamma_{xx}]_2^m)
X^{\b \sigma}([\gamma_{xx}]_{m+1}^q)
X^{\b \sigma'}([\gamma_{xx}]_{q+1}^r)
X^{\b \lambda_1}([\gamma_{xx}]_{r+1}^p\gamma^{(1)}{_x^o})
\label{ult}
\end{eqnarray}
Notice that the sets $\b \lambda_1$ and $\b \lambda_2$ will be joined by
a group product once this result is introduced into (\ref{aII3}). Each of
these terms generate then the following composition of loops

\begin{equation}
\DG_{\b \lambda_1 \b \lambda_2}^{\b \sigma_k}
X^{\b \lambda_1}([\gamma_{xx}]_{r+1}^p\gamma^{(1)}{_x^o})
X^{\b \lambda_2}(\gamma^{(1)}{_o^x} [\gamma_{xx}]_2^m)=
X^{\b \sigma_k}([\gamma_{xx}]_{r+1}^p\gamma^{(1)}{_x^o}
\gamma^{(1)}{_o^x} [\gamma_{xx}]_2^m)
\label{uult}
\end{equation}
We get from (\ref{aII3}), (\ref{ult}) and (\ref{uult})

\begin{eqnarray}
\lefteqn{ \epsilon_{abc}
X^{(ax\, \b \sigma_1\,bx\,\b \sigma_2 \,cx\,\b \sigma_3)_c}(\gamma_o) }
\nonumber\\
&&=\epsilon_{abc}\sum_{m,q,r}T_m^{ax}T_q^{bx}T_r^{cx} \,
X^{(\b \sigma_1\mid}(\gamma_{rm})
X^{\mid\b \sigma_2\mid}(\gamma_{mq})
X^{\mid\b \sigma_3)_c}(\gamma_{qr})
\end{eqnarray}
where $(\cdot\mid\mid\cdot\mid\mid\cdot)_c$ means the cyclic permutation
of the {\it sets} of indices. The other contribution of (\ref{aII4}) can
be developed in a similar way. In this case we have the reroutings
generated by the overline operation:

\begin{eqnarray}
&&\hspace{-0.4cm}\epsilon_{abc}
X^{(ax\, \bb \sigma_3\,bx\,\bb \sigma_2 \,cx\,\bb \sigma_1)_c}(\gamma_o)
\nonumber\\
&&=\epsilon_{abc}\sum_{m,q,r}T_m^{ax}T_q^{bx}T_r^{cx}
X^{(\bb \sigma_3\mid}(\gamma_{rm})
X^{\mid\bb \sigma_2\mid}(\gamma_{mq})
X^{\mid\bb \sigma_1)_c}(\gamma_{qr})\nonumber\\
&&=\epsilon_{abc}\sum_{m,q,r}T_m^{ax}T_q^{bx}T_r^{cx}\,
X^{(\b \sigma_3\mid}(\overline{\gamma}_{rm})
X^{\mid\b \sigma_2\mid}(\overline{\gamma}_{mq})
X^{\mid\b \sigma_1)_c}(\overline{\gamma}_{qr})
\end{eqnarray}
Using the above results one can write

\begin{eqnarray}
&&\epsilon_{abc}
R^{(ax\, \b \sigma_1\,bx\,\b \sigma_2 \,cx\,\b \sigma_3)_c}=
\textstyle{1\over2}\epsilon_{abc}
\sum_{m,q,r} T_m^{ax}T_q^{bx}T_r^{cx} \times\\
&&[X^{(\b \sigma_1\mid}(\gamma_{rm})
X^{\mid\b \sigma_2\mid}(\gamma_{mq})
X^{\mid\b \sigma_3)_c}(\gamma_{qr})+
X^{(\b \sigma_3\mid}(\overline{\gamma}_{rm})
X^{\mid\b \sigma_2\mid}(\overline{\gamma}_{mq})
X^{\mid\b \sigma_1)_c}(\overline{\gamma}_{qr})]\nonumber
\end{eqnarray}
Now it is straightforward to evaluate the r.h.s of (\ref{aII5}).
For the first contribution of (\ref{aII5}) we have
\begin{eqnarray}
\lefteqn{ \epsilon_{abc}
\DG_{\b \sigma_1 \b \sigma_2}^{\b \alpha}
R^{(ax\, \b \sigma_2\,cx\,\b \beta \,bx\,\bb \sigma_1)_c}(\gamma_o)
=\textstyle{1\over2}\epsilon_{abc}
\sum_{m,q,r} T_m^{ax}T_q^{bx}T_r^{cx} \times }
\nonumber\\
&&[X^{(\b \sigma_2\mid}(\gamma_{rm})
X^{\mid\b \beta\mid}(\gamma_{mq})
X^{\mid\bb \sigma_1)_c}(\gamma_{qr})+
X^{(\bb \sigma_1\mid}(\overline{\gamma}_{rm})
X^{\mid\b \beta \mid}(\overline{\gamma}_{mq})
X^{\mid\b \sigma_2)_c}(\overline{\gamma}_{qr})]
\nonumber\\
&&=\textstyle{1\over2}\epsilon_{abc}
\sum_{m,q,r}T_m^{ax}T_q^{bx}T_r^{cx} \times
\nonumber\\
&&[X^{\b \alpha}(\overline{\gamma}_{qr}\gamma_{rm})
X^{\b \beta}(\gamma_{mq})+
X^{\b \alpha}(\overline{\gamma}_{mq}\gamma_{qr})
X^{\b \beta}(\gamma_{rm})+
X^{\b \alpha}(\overline{\gamma}_{rm}\gamma_{mq})
X^{\b \beta}(\gamma_{qr})+
\nonumber\\
&&\hspace{0.1cm}X^{\b \alpha}(\gamma_{rm}\overline{\gamma}_{qr})
X^{\b \beta}(\overline{\gamma}_{mq})+
X^{\b \alpha}(\gamma_{qr}\overline{\gamma}_{mq})
X^{\b \beta}(\overline{\gamma}_{rm})+
X^{\b \alpha}(\gamma_{mq}\overline{\gamma}_{rm})
X^{\b \beta}(\overline{\gamma}_{qr})]\nonumber\\
\label{aII6}
\end{eqnarray}
A similar result is obtained for the other contribution:

\begin{eqnarray}
&&\epsilon_{abc}\DG_{\b \sigma_1 \b \sigma_2}^{\b \beta}
R^{(ax\, \b \alpha\,bx\,\b \sigma_1 \,cx\,\bb \sigma_2)_c}(\gamma_o)=
\textstyle{1\over2}\epsilon_{abc}
\sum_{m,q,r}T_m^{ax}T_q^{bx}T_r^{cx} \times
\nonumber\\
&&[X^{\b \alpha}(\gamma_{mq})
X^{\b \beta}(\gamma_{qr}\overline{\gamma}_{rm})+
X^{\b \alpha}(\gamma_{qr})
X^{\b \beta}(\gamma_{rm}\overline{\gamma}_{mq})+
X^{\b \alpha}(\gamma_{rm})
X^{\b \beta}(\gamma_{mq}\overline{\gamma}_{qr})+
\nonumber\\
&&\hspace{0.1cm}X^{\b \alpha}(\overline{\gamma}_{mq})
X^{\b \beta}(\overline{\gamma}_{rm}\gamma_{qr})+
X^{\b \alpha}(\overline{\gamma}_{qr})
X^{\b \beta}(\overline{\gamma}_{mq}\gamma_{rm})+
X^{\b \alpha}(\overline{\gamma}_{rm})
X^{\b \beta}(\overline{\gamma}_{qr}\gamma_{mq})]
\nonumber\\
\label{aII7}
\end{eqnarray}
Introducing now (\ref{aII6}) and (\ref{aII7}) into (\ref{aII5}) we get
the result (\ref{aII0}).

\end{document}